# Electronic and elastic properties of new nitrogen-containing perovskite-like superconductor ZnNNi$_3$

I.R. Shein*, V.V. Bannikov, A.L. Ivanovskii

*Institute of Solid State Chemistry, Ural Branch of the Russian Academy of Sciences, 620041, Ekaterinburg, Russia*

**Abstract**

Full-potential linearized augmented plane wave (FP-LAPW) method with the generalized gradient approximation (GGA) for the exchange-correlation potential has been applied for the study of structural, elastic and electronic properties of the newly synthesized nitrogen-containing perovskite-like superconductor ZnNNi$_3$. The optimized lattice parameter, independent elastic constants ($C_{11}$, $C_{12}$ and $C_{44}$), bulk modulus $B$, compressibility $\beta$, and shear modulus $G$ are evaluated. The band structure, total and site- projected $l$- decomposed DOSs, the shape of the Fermi surface, the Sommerfeld coefficient and the molar Pauli paramagnetic susceptibility for this novel anti-perovskite are obtained and analyzed – in comparison with related anti-perovskites ZnCNi$_3$ and MgCNi$_3$

*E-mail address:* shein@ihim.uran.ru (I.R. Shein).

# 1. Introduction

Since the discovery of 8-K superconductivity for cubic anti-perovskite MgCNi$_3$ [1] a much interest have been paid to others related materials, and the researches of a set of ternary carbon-containing anti-perovskites $M$CNi$_3$ have been intensively performed. So, ZnCNi$_3$, AlCNi$_3$, GaCNi$_3$, as well as some others $M^{II}$CNi$_3$ or $M^{III}$CNi$_3$ anti-perovskites, where $M$ are di- ($M^{II}$) or trivalent ($M^{III}$) cations have been synthesized and some of their physical properties have been investigated [2-16].

For the anti-perovskites $M^{II}$CNi$_3$, the interest is mainly due to the preference of superconductivity over magnetism which is rather unexpected in such Ni-rich compounds. Though the origin of the superconductivity in $M^{II}$CNi$_3$ is still controversial and remains an open problem; the available data (see reviews [6,7]) gives some evidences of conventional s-wave BCS type behavior. The significant achievements in the fundamental understanding of the coupling mechanism and electronic properties of these superconducting $M^{II}$CNi$_3$ materials were provided by means of the first principle band structure calculations [7-16]. It was shown that the density of states (DOS) at the Fermi level, $N(E_F)$, is dominated by Ni $d$ states with admixture of C $2p$ states forming π-type bands. Besides it has been pointed out [6,7] that the superconducting $M^{II}$CNi$_3$ compounds tend toward the ferromagnetism due to the existence of a van Hove singularity in the DOS slightly below the Fermi level. As this peak is sharp and the Fermi level lies on the shoulder of the peak, hole doping causes an increase in $N(E_F)$, resulting in a ferromagnetic instability and suppression of the superconducting transition. On the other hand, the electron doping causes a rapid decrease in $N(E_F)$, resulting also in a transition temperature suppression. Really, the anti-perovskites with trivalent metals $M^{III}$ (for example AlCNi$_3$ or GaCNi$_3$) are non-superconducting.

In this context, the recent discovery of a new superconducting ($T_C$~3K) nitrogen-containing anti-perovskite ZnNNi$_3$, which has been successfully prepared by reacting Zn with Ni powders in NH$_3$ gas [17], is very intriguing. Really, the related carbon-containing anti-perovskite ZnCNi$_3$ remains a paramagnetic metal up to T < 2 K [2]. Thus, in framework of the rigid-band picture, the ZnNNi$_3$ anti-perovskite (which is isoelectronic with above mentioned non-superconducting $M^{III}$CNi$_3$ phases) can be considered as one-electron doped ZnCNi$_3$ compound, where the Fermi level should be located far from the Ni 3$d$ peak, *i.e.* in the region of a quite low density of states that is unfavorable for superconductivity.

In this Communication, we report a detailed study by means of the FLAPW method within the generalized gradient approximation (GGA) of the structural, elastic and electronic properties of newly synthesized non-oxide anti-perovskite ZnNNi$_3$ (as far as we know, ZnNNi$_3$ is the first nitrogen-containing superconducting material in Ni-based anti-perovskite series) - in comparison with related carbon-containing anti-perovskite ZnCNi$_3$.

## 2. Computational method

The considered ZnNNi$_3$ anti-perovskite adopt [17] cubic structure (s.g. P$m3m$) consisting of Zn ions at the corners, nitrogen at the body center, and Ni at the face centers of the cube. The atomic positions are Ni: 3$c$ (½,½,0); Zn: 1$a$ (0,0,0) and N: 1$b$ (½,½,½). The band-structure calculations were done by means of the full potential method with mixed basis APW+lo (LAPW) implemented in the WIEN2k suite of programs [18]. The generalized gradient correction (GGA) to exchange-correlation potential in PBE form [19] was used. The electronic configurations were taken to be [Ar] $3d^84s^24p^0$ for Ni, [Ar] $3d^{10}4s^24p^0$ for Zn, and [He] $2s^22p^3$ for nitrogen. Here, the noble gas cores are distinguished from the sub-shells of valence electrons. The basis set inside each *muffin-tin* (MT) sphere is split into core and valence subsets. The core states are treated within the spherical part of the potential only, and are assumed to have a spherically symmetric charge density confined within MT spheres. The valence part is treated with the potential expanded to spherical harmonics up to $l = 4$. The valence wave functions inside the spheres are expanded up to $l = 12$. The plane-wave expansion with $R_{MT} \times K_{MAX}$ equal to 8, and $k$ sampling with 12×12×12 $k$-points mesh in the Brillouin zone were used. Relativistic effects are taken into account within the scalar-relativistic approximation. The MT radii were 1.88, 2.50 and 1.66 a.u. for Ni, Zn and nitrogen, respectively.

The self-consistent calculations were considered to be converged when the difference in the total energy of the crystal did not exceed 0.01 mRy as calculated at consecutive steps. The density of states (DOS) was obtained using a modified tetrahedron method [20].

## 3. Discussion

Firstly, the equilibrium lattice constant ($a_0$) for the ideal stoichiometric ZnNNi$_3$ is calculated. The result obtained $a_0^{theor}$ =3.769 Å appear to be in a reasonable accordance with the experiment: $a_0^{exp}$ = 3.756 Å [17], and the reduction of the experimentally obtained lattice constant of ZnNNi$_3$ occurs probably from the deviation of the investigated samples [17] from the stoichiometry (nitrogen deficient samples). As compared with the same estimations for the stoichiometric ZnCNi$_3$ [16], $a_0$(ZnCNi$_3$ = 3.772 Å) > $a_0$(ZnNNi$_3$) and this result can be explained by the atomic radii of carbon and nitrogen: R(C) = 0.77 Å > R(N) = 0.71 Å.

Secondly, let us discuss the mechanical parameters for ZnNNi$_3$ single crystal as obtained within the framework of the FLAPW–GGA calculations. The values of elastic constants ($C_{ij}$) are presented in Table 1. These three independent elastic constants in a cubic symmetry ($C_{11}$, $C_{12}$ and $C_{44}$) were estimated by calculating the stress tensors on applying strains to an equilibrium structure. All $C_{ij}$ constants for ZnNNi$_3$ crystal are positive and satisfy the generalized criteria [21] for mechanically stable crystals: ($C_{11}$ - $C_{12}$) > 0; ($C_{11}$ + 2$C_{12}$) > 0; $C_{44}$ > 0. Such crystals are characterized by positive values of the bulk modulus $B = (C_{11}+ 2C_{12})/3$ and shear modulus $G = C_{44}$. According to our calculations $B > G$; this implies that the parameter limiting the mechanical stability of this material is the shear modulus $G$. According to the criterion [22], a material is brittle if the $B/G$ ratio is less than 1.75. In our case, for $ZnNNi_3$ this value is 6.26, this means that $ZnNNi_3$ will behave as a brittle material.

In turn, comparing $ZnNNi_3$ and $ZnCNi_3$ (Table 1) it can be seen, that the bulk modules $B$ decreases as going from $ZnNNi_3$ to $ZnCNi_3$, *i.e.* in reversible sequence to $a_0$ - in agreement with the well-known relationship [23] between $B$ and the lattice constants (cell volume $V_o$, as $B \sim V_o^{-1}$). Accordingly, the compressibility of these species changes as $\beta(ZnNNi_3) < \beta(ZnCNi_3)$.

Let us discuss the electronic properties for $ZnNNi_3$. The band structure and the Fermi surface which have been calculated for the equilibrium geometry of $ZnNNi_3$, are shown in Figs. 1 and 2, respectively. The valence region which extends from -8.4 eV up to the Fermi level $E_F = 0$ eV is composed from the 15 $Ni3d$ and three $N2p$ bands filled by 35 electrons (Fig. 1). The $N 2s$ states are situated in a region far from the Fermi level $E_F$ (at about -15 eV below the Fermi level, and are not shown in Fig. 1), whereas Zn states are localized in a narrow interval close to -7.1 eV. Both of these states play a relatively minor role in valence area. Let us also note that $N2p$ and $Ni3d$ derived bands are separated by gap at about 0.4 eV, Fig. 1.

In valence region, the $N 2p$ states are partially hybridized with $Ni 3d$, and located below the near-Fermi bands formed mainly from $Ni 3d$ states, see Fig. 3. From these bands three are bonding, three - nonbonding, and three - antibonding bands. For $ZnNNi_3$ the antibonding $\pi$-bands confined between -0.7 eV and 0.9 eV cross $E_F$ along the $\Gamma$-R, R-X and $\Gamma$-M lines. These bands produce the hole- and electron-like sheets of the Fermi surface (Fig. 2) This topology of the Fermi surface for $ZnNNi_3$ is similar to the Fermi surfaces of other superconducting $M^{II}CNi_3$ materials, see [6,7,15,16].

For $ZnNNi_3$ the Fermi level is located at the high-energy slope of the peak which is associated with the quasi-flat $Ni 3d$ band (as well as for superconducting $MgCNi_3$) but the DOS at the Fermi energy for $ZnNNi_3$ (Table 2) with respect to that of $MgCNi_3$ (5.280 states/eV cell) is reduced by about 2.813 states/eV cell, i.e. more than by about 53 %. This is a simple argument, why $T_C$ decreases from 8K (for $MgCNi_3$) up to 3K (for $ZnNNi_3$). On the other hand, the value of $N(E_F)$ for non-superconducting $ZnCNi_3$ is also higher than the value of $N(E_F)$ for $ZnNNi_3$, as well as the Sommerfeld constants ($\gamma$) and the Pauli paramagnetic susceptibility ($\chi$) for stoichiometric $ZnCNi_3$ anti-perovskite, which are estimated assuming the free electron model, as: $\gamma = (\pi^2/3)N(E_F)k^2_B$, and $\chi = \mu_B^2 N(E_F)$, Table 2.

Thus, from our calculations, ideal stoichiometric $ZnNNi_3$, $ZnCNi_3$ and $MgCNi_3$ are very alike in both structural and electronic properties, but $MgCNi_3$ and $ZnNNi_3$ are superconducting materials, whereas $ZnCNi_3$ is reported as non - superconducting metal. Probably, the origin of this discrepancy should be attributed to the effect of non-stoichiometry. Really, for $MgCNi_3$ the presence of the C-vacancies results in suppression of the superconducting transition, see []. It may be assumed that for $ZnNNi_3$ anti-perovskite (which can be considered as one-electron doped $MgCNi_3$) the presence of the vacancies in the nitrogen lattice can lead to a reverse effect. Thus in our opinion a detailed study of the influence of nitrogen vacancy on the electronic spectra of nitrogen-deficient $ZnN_{1-x}Ni_3$ is quite necessary to understand the nature of superconductivity in this interesting material.

**4. Conclusions**

In summary, the first principle FLAPW-GGA method has been used for study of the structural, elastic and electronic properties of the newly synthesized 3K superconductor $ZnNNi_3$ as the first nitrogen-containing superconducting material in Ni-based anti-perovskite series.

From our calculations, the stoichiometric $ZnNNi_3$ crystal is mechanically stable. For $ZnNNi_3$ $B > G$, *i.e.* the parameter limiting the mechanical stability of this material is the shear modulus $G$. It was found also that $ZnNNi_3$ will behave as a brittle material.

In turn, comparing $ZnNNi_3$ and others carbon-containing anti-perovskites $MCNi_3$ it can be seen, that the bulk modulus $B$ for $ZnNNi_3$ (~ 204.7 GPa) adopt the maximal value as compared with $ZnCNi_3$ (177.0 GPa), $MgCNi_3$ (171.1 GPa) [16], $CdCNi_3$ (188.3 GPa) and $InCNi_3$ (187.5 GPa) [15]. Since a strong correlation exists between the bulk modulus and hardness of materials [25], the higher hardness should be for $ZnNNi_3$.

On the other hand, our calculations clearly show that ideal stoichiometric $ZnNNi_3$ and related compounds: $ZnCNi_3$ and $MgCNi_3$ are very alike in both structural and electronic properties, but $MgCNi_3$ and $ZnNNi_3$ are superconducting materials, whereas $ZnCNi_3$ is reported as non - superconducting metal. We believe that the origin of this discrepancy should be attributed to the effect of non-stoichiometry, thus a detailed study of the influence of nitrogen vacancy on the electronic spectra of nitrogen-deficient $ZnN_{1-x}Ni_3$ is quite necessary to understand the nature of superconductivity in this interesting material.

**References**

[1] T. He, Q. Huang, A.P. Ramirez, Y. Wang, K.A. Regan, N. Rogado, M.A. Hayward, M.K. Haas, J.S. Slusky, K. Inumaru, H.W. Zandbergen, N.P. Ong, and R.J. Cava, Nature (London) **411**, 54 (2001)

[2] M.S. Park, J.S. Giim, S.H. Park, Y.W. Lee, S.I. Lee, and E.J. Choi. Supercond. Sci. Technol. **17**, 274 (2004).

[3] R.E. Schaak, M. Avdeev, W.-L. Lee, G. Lawes, H.W. Zandbergen, J.D. Jorgensen, N.P. Ong, A.P. Ramirez, and R.J. Cava, J. Solid State Chem. **177**, 1244 (2004).

[4] A.F. Dong, G.C. Che, W.W. Huang, S.L. Jia, H. Chen, and Z.X. Zhao, Physica C **422**, 65 (2005).

[5] P. Tong, Y.P. Sun, X.B. Zhu, and W.H. Song, Phys. Rev. B **73**, 245106 (2006).

[6] A.L. Ivanovskii, Phys. Solid State **45**, 1829 (2003).

[7] S. Mollah, J. Phys.: Condens. Matter **16**, R1237 (2004).
[8] M.D. Johannes, W.E. Pickett, Phys. Rev. B **70**, 060507 (2004).
[9] C.M.I. Okoye, Solid State Commun. **136**, 605 (2005).
[10] V.V. Bannikov, I.R. Shein, and A.L. Ivanovskii, Phys. Solid State **49**, 1626 (2007).
[11] T. Klimczuk, V. Gupta, G. Lawes, A.P. Ramirez, and R.J. Cava, Phys. Rev. B **70**, 094511 (2004).
[12] S.-H. Park, Y.W. Lee, J. Giim, S.-H. Jung, H.C. Ri, and E.J. Choi, Physica C **400**, 160 (2004).
[13] T. Klimczuk, M. Avdeev, J.D. Jorgensen, and R.J. Cava, Phys. Rev. B **71**, 184512 (2005).
[14] I.R. Shein, A.L. Ivanovskii, E.Z. Kurmaev, A. Moewes, S. Chiuzbian, L.D. Finkelstein, M. Neumann, Z.A. Ren, and G.C. Che. Phys. Rev. B **66**, 024520 (2002).
[15] I.R. Shein, and A.L. Ivanovskii, Phys. Rev., B **77**, 104101 (2008)
[16] I.R. Shein, V.V. Bannikov, and A.L. Ivanovskii, Physica C **468**, 1 (2008).
[17] M. Uehara, A. Uehara, K. Kozawa, and Y. Kimishima, arXiv/cond-mat.0811.3483 (2008).
[18] P. Blaha, K. Schwarz, G.K.H. Madsen, D. Kvasnicka, and J. Luitz, WIEN2k, *An Augmented Plane Wave Plus Local Orbitals Program for Calculating Crystal Properties,* (Vienna University of Technology, Vienna, 2001).
[19] J.P. Perdew, S. Burke, and M. Ernzerhof, Phys. Rev. Lett. **77**, 3865 (1996).
[20] P.E. Blochl, O. Jepsen, and O.K. Anderson, Phys. Rev. B **49**, 16223 (1994).
[21] G. Grimvall, *Thermophysical Properties of Materials* (North-Holland, Amsterdam, 1986).
[22] S.F. Pugh, Phil. Mag. **45**, 833 (1954).
[23] M.L. Cohen. Phys. Rev. B **32**, 7988 (1985).
[24] P.J.T. Joseph and P.P. Singh, J. Phys.: Condens. Matter **18,** 5333 (2006).
[25] J. Haines, J.M. Leger, and G. Bocquillon, Ann. Rev. Mater. Res. **31**, 1 (2001).

**Table 1.** Calculated lattice parameter ($a_o$, in Å), elastic constants ($C_{ij}$, in GPa), bulk modulus ($B$, in GPa), compressibility ($\beta$, in GPa$^{-1}$) and shear modules ($G = C_{44}$, in GPa) for cubic anti-perovskite ZnNNi$_3$ in comparison with ZnCNi$_3$ [16].

| Parameters | ZnNNi$_3$ | ZnCNi$_3$ |
|---|---|---|
| $a_o$ | 3.769 | 3.772 |
| $C_{11}$ | 364.20 | 319.53 |
| $C_{12}$ | 124.90 | 105.72 |
| $C_{44}$ ($G$) | 32.69 | 39.42 |
| $B$ | 204.66 | 176.99 |
| $\beta$ | 0.004886 | 0.005650 |

**Table 2.** Total and site- projected $l$- decomposed densities of states at the Fermi level (N(E$_F$), states/eV cell), calculated Sommerfeld coefficient ($\gamma$) and molar Pauli paramagnetic susceptibility ($\chi$) for cubic anti-perovskite ZnNNi$_3$ - in comparison with ZnCNi$_3$ [16].

| Parameters * | ZnNNi$_3$ | ZnCNi$_3$ |
|---|---|---|
| N(E$_F$)(C,N2$s$) | 0.01 | 0.019 |
| N(E$_F$)(C,N2$p$) | 0.328 | 0.323 |
| N(E$_F$)(Ni4$s$) | 0.025 | 0.057 |
| N(E$_F$)(Ni4$p$) | 0.044 | 0.087 |
| N(E$_F$)(Ni4$d$) | 2.117 | 3.141 |
| **N(E$_F$)(Ni)** | **2.187** | **3.285** |
| N(E$_F$)(ZM $s$) | 0.012 | 0.058 |
| N(E$_F$)(M $p$) | 0.031 | 0.058 |
| N(E$_F$)(M $d$) | 0.005 | 0.009 |
| **N(E$_F$) (total)** | **2.813** | **4.341** |
| $\gamma$, in mJ/mol f.u. K$^2$ | 6.631 | 10.23 |
| $\gamma_{Ni}$, in mJ/mol Ni K$^2$ | 5.155 | 7.745 |
| $\chi$, in 10$^{-4}$ emu/mol | 0.905 | 1.40 |

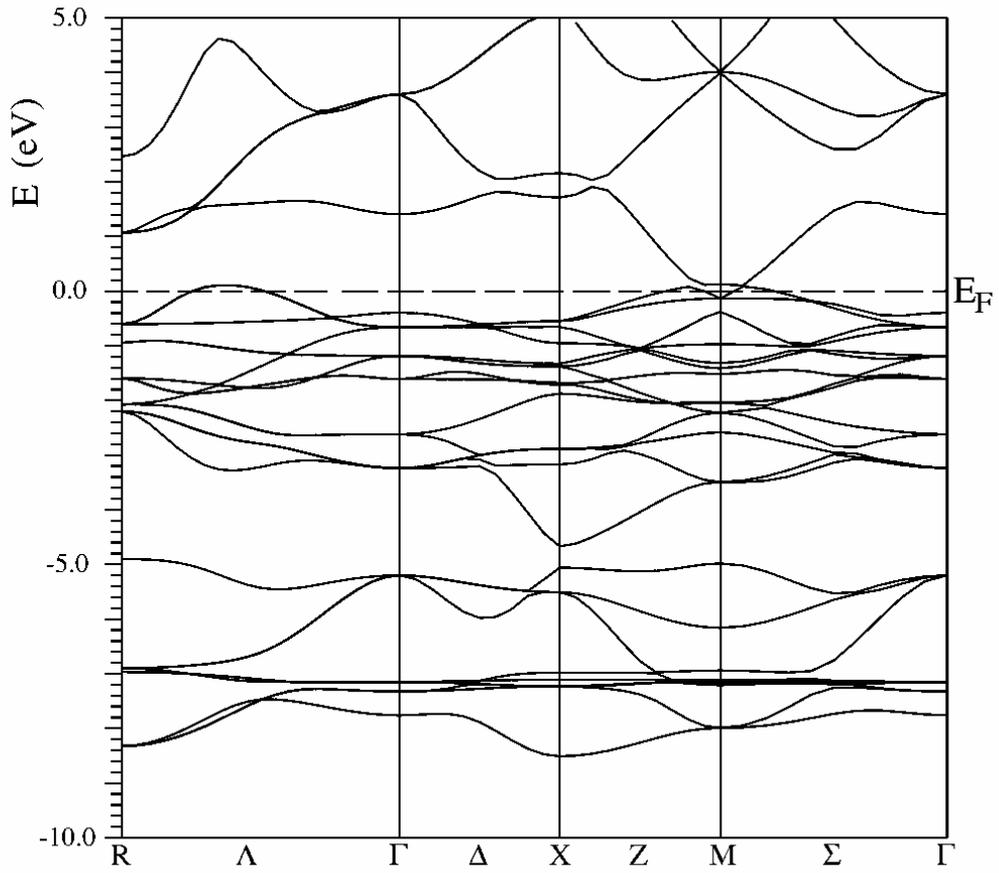

Fig. 1. Band structure of ZnNNi$_3$ along the symmetry lines of the cubic BZ. Fermi level $E_F$ = 0 eV.

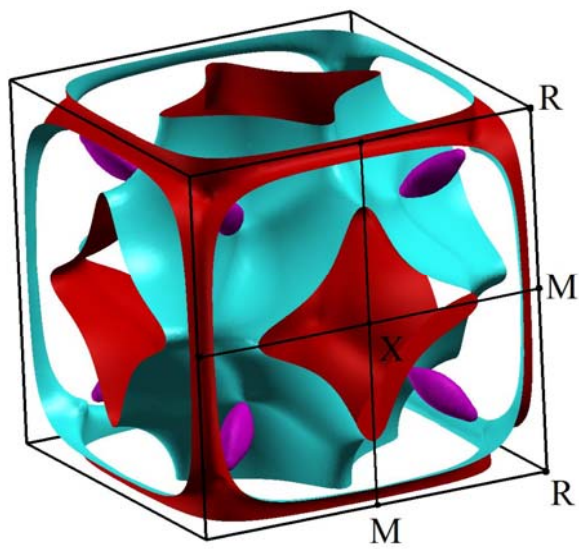

Fig. 2. The Fermi surfaces of ZnNNi$_3$.

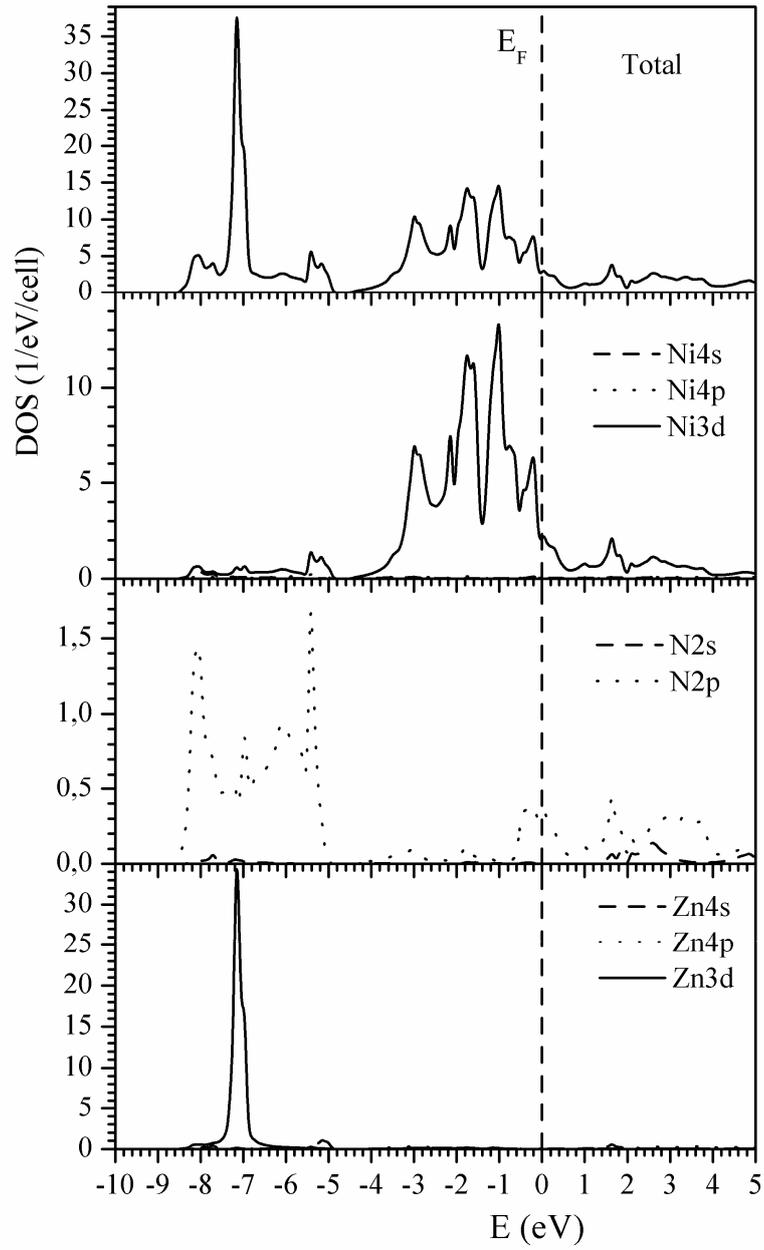

Fig. 3. Total and site- projected *l*- decomposed densities of states for ZnNNi$_3$.